\documentclass{article}
\usepackage{spconf,amsmath,epsfig}

\usepackage{graphicx}                 
\usepackage{amssymb,amsfonts} 
\usepackage{array}
\usepackage{algorithmic}
\title{A Coded Bit-Loading Linear Precoded Discrete Multitone Solution for Power Line Communication}
\name{Fahad Syed Muhammmad*, Jean-Yves Baudais, Jean-Fran\c cois H\'elard, and Matthieu Crussi\`ere\thanks{*Email: fahad.syed-muhammad@ens.insa-rennes.fr}}
\address{Institute of Electronics and Telecommunications of Rennes, 35043 Rennes Cedex, France}
\begin{document}
%
\maketitle
\begin{abstract}
Linear precoded discrete multitone modulation (LP-DMT) system has been already proved advantageous with adaptive resource allocation algorithm in a power line communication (PLC) context. In this paper, we investigate the bit and energy allocation algorithm of an adaptive LP-DMT system taking into account the channel coding scheme. A coded adaptive LP-DMT system is presented in the PLC context with a loading algorithm which accommodates the channel coding gains in bit and energy calculations. The performance of a concatenated channel coding scheme, consisting of an inner Wei's 4-dimensional 16-states trellis code and an outer Reed-Solomon code, in combination with the proposed algorithm is analyzed. Simulation results are presented for a fixed target bit error rate in a multicarrier scenario under power spectral density constraint. Using a multipath model of PLC channel, it is shown that the proposed coded adaptive LP-DMT system performs better than classical coded discrete multitone.
\end{abstract}
\section{Introduction}
The magic of discrete multitone modulation (DMT) is hidden in its orthogonal subcarriers, that enhance its capability to cope efficiently with narrowband interference, high frequency attenuations and multipath fadings with the help of simple equalization filters. Adaptive linear precoded discrete multitone (LP-DMT) system is based on classical DMT, combined with a linear precoding component. Combinations of multicarrier (MC) and linear precoding (LP) have proved their significance in the digital subscriber line (DSL) context~\cite{mallier}. Uncoded LP-DMT has already been suggested for power line communication (PLC) networks in~\cite{matthieu2} with a loading algorithm that handles the subcarrier, code, bit, and energy resource distribution among the active users but without taking into account the channel coding scheme. Assuming perfect channel state information (CSI) at the transmitting side, energy and bits are efficiently distributed among the precoding sequences by the loading algorithm to achieve either high throughput or high robustness. In this paper, we examine the performance of an LP-DMT system exploiting a resource allocation algorithm which takes into account the channel coding scheme. The loading algorithm, proposed in~\cite{matthieu2}, is modified to accommodate the coding gains associated with the channel coding scheme. The proposed bit and energy allocation algorithm can be used in combination with any channel coding scheme, no matter it has constant or variable coding gains for different modulation orders, provided the obtained coding gains are known for all the modulation orders. 

The suitable coding scheme for PLC networks should have large coding gains, reasonable implementation complexity and some measure of burst immunity. Selected on these bases, the proposed concatenated channel coding scheme consists of an inner Wei's 4-dimensional (4D) 16-states trellis code~\cite{wei} and an outer Reed-Solomon (RS) code. This combination has already proved its significance in xDSL systems and has been included in many standards~\cite{ituvdsl2}. The efficient performance of Wei's 4D 16-states trellis code over gaussian channel has also been demonstrated in~\cite{transcioffi}. Here, we apply a resource allocation algorithm to an LP-DMT system which takes into account this concatenated channel coding scheme.

The rest of the paper is organized as follows. In Section \ref{sysmod} the structure of a coded LP-DMT system is described. The modified bit loading algorithm is described in Section \ref{modalgo}. Section \ref{theory} gives the expression of the obtained coding gain along with the loss incurred due to redundancies. In Section \ref{simres}, simulation scenarios are discussed and results are presented for the classical DMT system and the proposed adaptive LP-DMT system using a multipath PLC channel model~\cite{zimmermann} for both coded and uncoded implementations. LP-DMT and DMT simulations, for both coded and uncoded scenarios, are run for various channel gains, while using the power line channel model suggested in~\cite{zimmermann}. It is shown that the proposed coded LP-DMT system with the modified bit loading algorithm performs better than coded DMT and achieves higher throughput for PLC applications. Finally, Section~\ref{conc} concludes this paper.
\section{System Descriptions}\label{sysmod}
\begin{figure}[!t]
\begin{center}
\leavevmode
\includegraphics[scale=0.217]{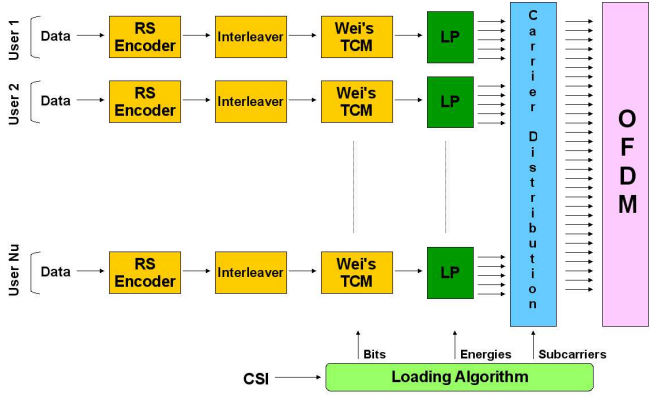} 
\caption{Coded LP-DMT transmitter structure}
\label{fig2}
\end{center}
\end{figure}

The structure of the suggested system, including the proposed channel coding scheme, is shown in Fig.~\ref{fig2}. The entire bandwidth is divided into $N$ parallel subcarriers which are split up into $N_k$ sets `$S_{k}$' of $L_c$ subcarriers. The precoding function is then applied block-wise by mean of precoding sequences of length $L_c$. In fact, the precoding function can be viewed as a spreading component carried out in the frequency domain as in multicarrier code-division-multiple-access (MC-CDMA). Factor $L_c$ is such that $L_c \ll N$, which implies that $N_{k} = \lfloor\frac{N}{L_{c}}\rfloor$. Note that the subsets in a given set are not necessarily adjacent. Each user $u$ of the network is being assigned a set $B_{u}$ of subsets $S_{k}$. We emphasize that $\forall{u},~B_{u}$ are mutually exclusive subsets. Consequently, multiple access between the $N_{u}$ users is managed following a frequency division multiple access (FDMA) approach, instead of a code division multiple access (CDMA) approach. It is worthy to mention here that we are going to consider the only case of a single user multiple block system which can be easily extended for a multi user multiple block scenario. The number of precoding sequences used to spread information symbols on one subset $S_{k}$ is denoted by $N_{c}^{(k)}$, with $0 \leq N_{c}^{(k)} \leq L_c$ since we assume orthogonal sequences. A certain amount of energy $e_{i}^{(k)}$ is assigned to each precoding sequence $c_{i}$ associated to a given modulation symbol of $b_{i}^{(k)}$ bits, where $1 \leq i \leq N_{c}^{(k)}$. Finally, in the following, $H = \{1,\ldots,N\}$ will be the set of the useful subcarriers of the multicarrier spectrum.

Wei's trellis code operates on the bits allocated to the precoding sequences and produces two 2D points at its output.
In Fig.~\ref{fig2}, only a single output is shown for Wei's trellis encoder, because both 2D outputs are allocated to the same precoding sequence. Also multiple copies of Wei's encoder is shown for the purpose of illustration whereas, in practice a single encoder is used to encode across the precoding sequences as discussed in~\cite{transcioffi}, where a single encoder is used to encode across the subcarriers. It will be shown in Section \ref{simres} that similar to independent and memoryless subchannels in a DMT scenario, precoding sequences are also independent and memoryless in an LP-DMT scenario. The gain obtained from the application of trellis code will therefore be the same as that obtained in an intersymbol interference (ISI) free environment.
The RS code operates on the binary stream at the input of the system, before the bit allocation block as shown in Fig.~\ref{fig2}. The RS code used here is a shortened RS code RS(240,224), supported in many standards~\cite{etsivdsl}, and can correct up to 8 erroneous bytes. A convolutional interleaver is used to spread the errors over a number of RS codewords.
\section{The Loading Algorithm}\label{modalgo}
The presented single user multiple block case of the bit loading algorithm takes into account the channel coding scheme and is modified to accommodate the channel coding gains in bit and energy calculations. Over a given subset of subcarriers $S_k$, the optimal achieved throughput under PSD constraint is given as
\begin{equation}\label{eqn-1}
R_{k} = L_c \log_2 \left( 1 + \frac{1}{\Gamma} \frac{L_c}{\sum\limits_{n\subset S_k} \frac{1}{|h_n|^2}} \frac{E_s}{N_0}\right)
\end{equation}
where $|h_n|^2$ is the gain of subchannel $n$, $\Gamma$ is defined as normalized signal to noise ratio (also known as SNR gap), $E_s$ is obtained from a given power spectral density (PSD) constraint and $N_0$ is the additive background white Gaussian noise level. SNR gap, $\Gamma$, has a constant value for all the modulation orders of uncoded quadrature amplitude modulation (QAM) for a fixed target SER. However, (\ref{eqn-1}) does not provide any practical throughput because it assumes infinite granularity.\footnote{infinite granularity: non-integer modulation order} The task here is therefore to find an appropriate bit distribution for $L_c$ available precoding sequences of each subset $S_k$, which maximizes the data rate. In the following, we discuss the allocation policy that handles the finite granularity problem.

Given an SNR gap $\Gamma$, a precoding factor $L_c$ and a transmission level $E_s$, the rate achieved by an adaptive LP-DMT system using discrete modulation is maximized if, on each subset $S_k$, $b_i^{(k)}$ bits are allocated to precoding sequence $c_i$ and $b_i^{(k)}$ is given as
\begin{equation}\label{bik}
b_i^{(k)} = 
\begin{cases} \lfloor R_k/L_c \rfloor +1 & (1 \leq i \leq n_c^{(k)})
\\
\lfloor R_k/L_c \rfloor & (n_c^{(k)} < i \leq N_c^{(k)})
\end{cases}
\end{equation}
where
\begin{equation*}
n_c^{(k)} = \lfloor L_c(2^{R_k/L_c-\lfloor R_k/L_c\rfloor} - 1)\rfloor.
\end{equation*}
Then, practically achievable data rate $\bar{R_k}$, considering finite granularity, on each subset $S_k$ is given as
\begin{align}
\bar{R_k}=&\lfloor L_c(2^{R_k/L_c-\lfloor R_k/L_c\rfloor} - 1)\rfloor \times (\lfloor R_k/L_c\rfloor + 1) \notag\\
&+ (L_c - \lfloor L_c(2^{R_k/L_c-\lfloor R_k/L_c\rfloor} - 1)\rfloor) \times \lfloor R_k/L_c\rfloor
\end{align}
Now we can optimally assign a particular modulation order to each precoding sequence on different subsets. Each precoding sequence's energy contribution in the total energy of a subcarrier of a given subset, $e_i^{(k)}$, is given as
\begin{equation}\label{eik}
e_i^{(k)} = (2^{b_i^{(k)}} - 1)\frac{\Gamma}{L_c^2}N_0 \sum_{n\subset S_k} \frac{1}{|h_n|^{2}}
\end{equation}

which satisfies $\displaystyle\sum_{i} e_i^{(k)}<E_s$. As discussed above, $\Gamma$ is defined for a given target SER with uncoded QAM and have a constant value for all the modulation orders. In this paper we are going to deal with fixed target bit error rate (BER) instead of SER. Then taking into account the coding gains and fixed BERs, $\Gamma$ is no more constant for all the modulation orders. The above algorithm is modified to accommodate variable $\Gamma$ for different modulation orders. The exact values of the SNR gaps for all the modulation orders are stored in a predefined table and are denoted by $\Gamma_{i}^{(k)}$. These values are calculated on the basis of the selected channel coding scheme and the required system margin, which we are going to discuss in Section~\ref{theory}. For a given subset $S_k$, initially, we can take any value for $\Gamma$, say $\Gamma_{i}^{init(k)}=1$. The closer the initial value to the exact value, the more efficient is the algorithm. $R_k$ is calculated from (\ref{eqn-1}) using $\Gamma_{i}^{init(k)}$ while $b_i^{(k)}$ and $n_c^{(k)}$ from (\ref{bik}). Now the exact value of SNR gap, $\Gamma_{i}^{(k)}$, is taken from the table depending upon the bit vector $b_i^{(k)}$, and $e_i^{(k)}$ is calculated from (\ref{eik}) using $\Gamma_{i}^{(k)}$. Gradually bits are added in the bit vector, $b_i^{(k)}$, till $\displaystyle\sum_i e_i^{(k)}$ exceeds the PSD limit, $E_s$, and subsequently bits are removed to respect the PSD limit. We can summarize the modified approach as follows:
\begin{algorithmic}[1]
\STATE Calculate $R_k$, $n_c^{(k)}$ and $b_i^{(k)}$ for $\Gamma_{i}^{init(k)}$
\STATE Now take $\Gamma_{i}^{(k)}$, depending upon $b_i^{(k)}$
\STATE Calculate $e_i^{(k)}$ for $\Gamma_{i}^{(k)}$
\STATE Start a counter, say $cnt=1$
\WHILE{$\left(\displaystyle\sum_i e_i^{(k)}>E_s\right)$}
\STATE $b_{(n_c+cnt)}^{(k)} = b_{(n_c+cnt)}^{(k)} + 1$
\STATE $cnt = cnt + 1$
\STATE update $e_i^{(k)}$
\ENDWHILE
\WHILE{$\left(\displaystyle\sum_i e_i^{(k)}>E_s\right)$}
\STATE $cnt = cnt - 1$
\STATE $b_{(n_c+cnt)}^{(k)} = b_{(n_c+cnt)}^{(k)} - 1$
\STATE update $e_i^{(k)}$
\ENDWHILE
\end{algorithmic}
\section{Theoretical coding gains}\label{theory}
In this section, the theoretical expressions are given for the selected channel coding scheme by using the assumptions given in~\cite{transcioffi}. The SNR gap $\Gamma$ for a target BER of $10^{-7}$ is given as
\begin{equation}\label{eqn-3}
\Gamma = 9.8 + \gamma_{m} - \gamma_{c}~~~~(dB)
\end{equation}
where $\gamma_{m}$ is the desired margin in the system and $\gamma_{c}$, the coding gain for the proposed concatenated channel coding scheme, is given as
\begin{equation}\label{eqn-4}
\gamma_{c} = \gamma_{tc,dB} + \gamma_{rs,dB} - \gamma_{loss,dB}~~~~(dB)
\end{equation}
where $\gamma_{tc,dB}$ and $\gamma_{rs,dB}$ are the gains provided by the trellis code and the RS code respectively and $\gamma_{loss,dB}$ is the loss incurred for increasing the data rate.\\
$\gamma_{rs,dB}$: The gain obtained by RS decoder is given as
\begin{equation}\label{eqn-9}
\gamma_{rs} = \Gamma_{0,P_{bit}} - \Gamma_{rs}~~~~(dB)
\end{equation}
where $\Gamma_{rs}$ is the SNR gap to obtain $P_b$, the BER at the output of the demodulator, and is given as\footnote{division by 2 in $Q^{-1}[.]$ results from the assumption that BER is one-half the 2-D error rate~\cite{transcioffi}.}
\begin{equation}\label{eqn-7}
\Gamma_{rs} = \frac{1}{3} \Bigg(Q^{-1}\bigg[\frac{P_{b}}{2}\bigg]\Bigg)^2
\end{equation}
and $\Gamma_{0,P_{bit}}$ is the SNR gap to obtain $P_{bit}$, the BER of the uncoded system, and is given as
\begin{equation}\label{eqn-8}
\Gamma_{0,P_{bit}} = \frac{1}{3} \Bigg(Q^{-1}\bigg[\frac{P_{bit}}{2}\bigg]\Bigg)^2
\end{equation}
$\gamma_{tc,dB}$: As $P_b$ is the required BER at the input to the RS decoder and $\Gamma_{0,P_{b}}$ and $\Gamma_{tc,P_{b}}$ are the SNR gaps required by an uncoded and a Wei's 4D 16-states trellis coded system respectively to achieve $P_{b}$. Then the coding gain of a Wei's 4D 16-states trellis code can be given by
\begin{equation}\label{eqn-10}
\gamma_{tc} = \Gamma_{0,P_{b}} - \Gamma_{tc,P_{b}}~~~~(dB)
\end{equation}
$\gamma_{loss,dB}$: If $P_{tot}^*(b)$ is the minimum amount of power required to achieve the data rate $b$ as defined in~\cite{transcioffi}, the loss for the increased data rate associated with the RS code, $\gamma_{loss,dB}$, can be given as
\begin{equation}\label{eqn-12}
\gamma_{loss,dB} = P_{tot,dB}^*(\frac{nb}{k}) - P_{tot,dB}^*(b)
\end{equation}
\section{Simulation results}\label{simres}
In this section, we will present simulation results for the proposed concatenated channel coding scheme combined with the adaptive LP-DMT system and the modified bit loading algorithm. The performance of the coded adaptive LP-DMT system is compared with that of the coded DMT, i.e., when $L_c$=1, and uncoded LP-DMT. 
\begin{figure}[!t]
\begin{center}
\includegraphics[scale=0.457]{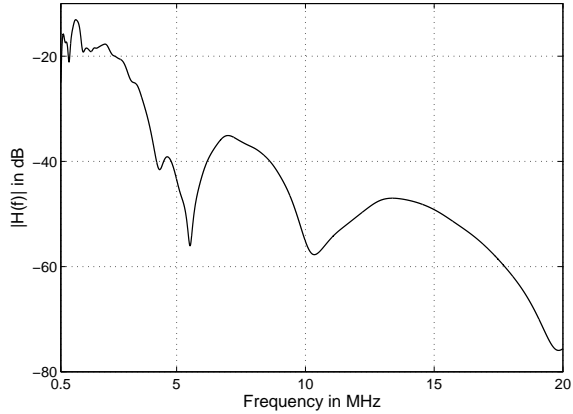} 
\caption{15-paths reference channel model for PLC~\cite{zimmermann}}
\label{fig3}
\end{center}
\end{figure}
\begin{table}[!t]
\caption{PARAMETERS OF THE 15-PATH MODEL}
\label{table1}
\begin{center}
\begin{tabular}{c|c|c|c|c|c}
\multicolumn{6}{c}{} \\
\hline\hline
\multicolumn{6}{c}{\textbf{attenuation parameters}} \\
\hline
\multicolumn{2}{c|}{$k=1$} & \multicolumn{2}{c|}{$a_0=0$} & \multicolumn{2}{c}{$a_1=2.5\cdot10^{-9}$} \\
\hline\hline
\multicolumn{6}{c}{\textbf{path-parameters}} \\
\hline
$i$ & $g_i$ & $d_i (m)$ & $i$ & $g_i$ & $d_i (m)$ \\
\hline
1 & 0.029 & 90 & 9 & 0.071 & 411 \\
\hline
2 & 0.043 & 102 & 10 & -0.035 & 490 \\
\hline
3 & 0.103 & 113 & 11 & 0.065 & 567 \\
\hline
4 & -0.058 & 143 & 12 & -0.055 & 740 \\
\hline
5 & -0.045 & 148 & 13 & 0.042 & 960 \\
\hline
6 & -0.040 & 200 & 14 & -0.059 & 1130 \\
\hline
7 & 0.038 & 260 & 15 & 0.049 & 1250 \\
\hline
8 & -0.038 & 322 & & & \\
\hline\hline
\end{tabular}
\end{center}
\end{table} 
The generated LP-DMT signal is composed of $N$=1024 subcarriers transmitted in the band [500;20,000]~kHz. The optimal value of the precoding factor is obtained by running the simulations for various possible values of $L_c$, which came out to be 32 for  said channel model. The subcarrier spacing is $19.043$~kHz. It is assumed that the synchronization and channel estimation tasks have been successfully performed. We use the multipath model for the power line channel as proposed in~\cite{zimmermann} and shown in Fig.~\ref{fig3}. The considered reference model is 110~m link 15-paths model whose frequency response is given by
\begin{equation}\label{eqn-11}
H(f) = \sum_{i=1}^{N} g_i \cdot e^{-(a_0+a_1f^k)^{d_i}} \cdot e^{-j2\pi{f(\tau_i)}}
\end{equation}
a result which has been widely proved in practice. The parameters of the 15-path model are listed in Table~\ref{table1}, and $\tau_i$ is the delay of path $i$. A background noise level of -110 dBm/Hz is assumed and the signal is transmitted with respect to a flat PSD of -40 dBm/Hz. The maximum number of bits per symbol is limited to 10 and minimum number of bits per symbol is 2. Results are given for a fixed target BER of $10^{-7}$.

\begin{figure}[!t]
\begin{center}
\includegraphics[scale=0.457]{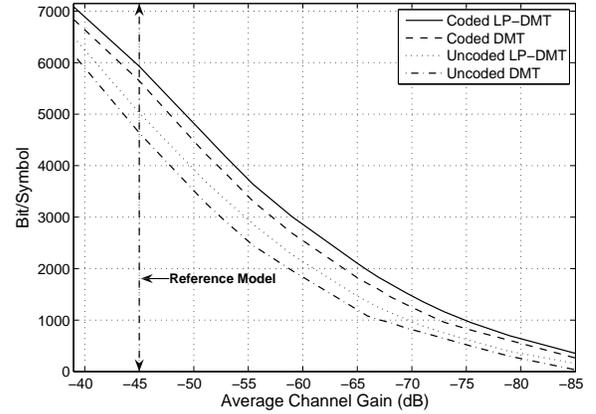} 
\caption{Achieved throughputs at various channel gains}
\label{fig4}
\end{center}
\end{figure}

Fig.~\ref{fig4} shows the achieved bit per DMT symbol versus the average channel gain $G=(\frac{1}{N})\sum{|h_n|^2}$ which conveys the attenuation experienced by the signal through the channel. The corresponding SNR is then given by $SNR=-30+G_{dB}+110$. The performance of LP-DMT is compared with DMT at different average channel gains for both coded and uncoded implementations, where coded LP-DMT implementation uses the proposed algorithm as discussed in Section \ref{modalgo}. The DMT system can be obtained by taking $L_c=1$ in the LP-DMT system. The simulations are run for a single user multiple block scenario. The proposed adaptive coded LP-DMT system can easily be extended to a multi user multiple block scenario. 
\begin{figure}[!t]
\begin{center}
\includegraphics[scale=0.457]{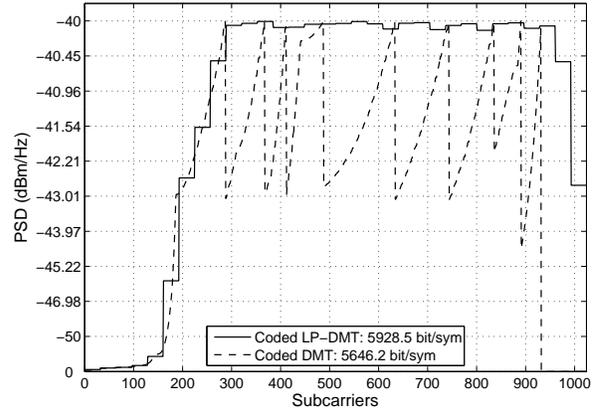} 
\caption{Energy distribution comparison}
\label{fig6}
\end{center}
\end{figure}
The reason for the better performance of the coded LP-DMT system is explained in Fig.~\ref{fig6}, where energy distribution of the coded LP-DMT is compared with that of the coded DMT. The spike-shaped curve of the coded DMT shows the transitions of the modulation orders (i.e. decreasing the constellation sizes) when no more energy is available to sustain the fixed target BER. It is clear that the coded DMT is not fully exploiting the available energy on each subcarrier due to finite granularity and PSD constraints, while the precoding component of the coded LP-DMT system accumulates the energies of a given subset of subcarriers to transmit additional bits. Both systems respect the PSD constraint of -40 dBm/Hz as defined earlier. The coded adaptive LP-DMT system utilizes more efficiently this PSD limit in comparison with the coded DMT system. Fig.~\ref{fig6} gives the minimal required energy allowing the transmission of the maximum data rate. According to the PSD mask, the residual available energy would not lead to any increase in the data rate.
\begin{figure}[!t]
\begin{center}
\includegraphics[scale=0.457]{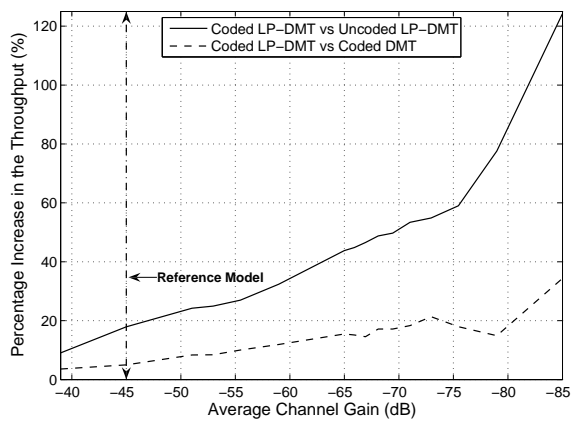} 
\caption{Percentage increase in the throughput}
\label{fig7}
\end{center}
\end{figure}

Fig.~\ref{fig7} gives the percentage increase in the throughput obtained by coded LP-DMT in comparison of coded DMT and uncoded LP-DMT for various values of average channel gain. The vertical broken line in Fig.~\ref{fig7} represents the reference power line channel model proposed by Zimmermann. As it is shown in Fig.~\ref{fig6} and~\ref{fig7} that coded LP-DMT has the highest throughput and more efficient energy utilization when compared to coded DMT (see Fig.~\ref{fig6}). Furthermore, it is worth noting that coded LP-DMT is all the more interesting when the channel gain is low, i.e., the reception SNR is low. For instance, at an average channel gain of -60 dB, there is an improvement in the throughput of approximately 34.3\% and 12.5\% when we compare our coded LP-DMT system with uncoded LP-DMT and coded DMT respectively for a fixed target BER of $10^{-7}$. This improvement becomes 51\% and 17.5\% respectively at an average channel gain of -70 dB and is even higher at lower channel gains. Hence, we conclude that the proposed system can especially be advantageously exploited for poor SNR which is equivalent to claim that the proposed system ensures reliable communication over longer links and effectively increases the range of PLC systems.
\section{Conclusion}\label{conc}
In this paper, the resource allocation problem of a coded adaptive LP-DMT system is investigated. The proposed resource allocation algorithm incorporates the effect of channel coding in energy and bit distribution and handles various values of the SNR gap depending upon the constellation size. The performance of the proposed system is compared with that of coded DMT and uncoded LP-DMT and it is shown that the proposed system was able to transmit higher rates especially for lower channel gains. The performance is compared while applying the same proposed coding scheme to both the systems, although the proposed algorithm can be used in combination with any channel coding scheme provided the SNR gaps of that scheme are known for all the modulation orders. It is shown that using a powerful but low-complexity coding scheme with the proposed algorithm, improves the throughput of the system significantly and can especially be advantageously exploited for poor SNR.




%




\end{document}